\documentclass[twocolumn,prd,showpacs,preprintnumbers,amsmath,amssymb,floatfix]{revtex4}

\usepackage{graphicx}% Include figure files
\usepackage{dcolumn}% Align table columns on decimal point
\usepackage{bm}% bold math

\newcommand{\bra}[1]{\langle \, #1 \, |}
\newcommand{\ket}[1]{| \, #1 \, \rangle}

\begin{document}

\preprint{}

\title{Phenomenology of spin 3/2 baryons with
pentaquarks}% Force line breaks with \\

\author{Tetsuo~Hyodo}
\email{hyodo@rcnp.osaka-u.ac.jp}
\affiliation{%
Research Center for Nuclear Physics (RCNP),
Ibaraki, Osaka 567-0047, Japan
}%

\author{Atsushi~Hosaka}%
\affiliation{%
Research Center for Nuclear Physics (RCNP),
Ibaraki, Osaka 567-0047, Japan
}%

\date{\today}% It is always \today, today,
\begin{abstract}
    We examine several assignments of spin and parity for the
    pentaquark $\Theta^+$ state ($J^P=1/2^{\pm}, 3/2^{\pm}$)
    in connection with known baryon resonances.
    Assuming that the $\Theta^+$ belongs to an antidecuplet representation
    which mixes with an octet,
    we calculate the mass spectra of the flavor partners of 
    the $\Theta^+$
    based on the SU(3) symmetry.
    The decay widths of the $\Theta^+$
    and nucleon partners are analyzed for 
    the consistency check of the mixing angle
    obtained from the masses.
    It is found that
    a suitable choice of the mixing angle successfully
    reproduces the observed masses of 
    $\Theta(1540)$ and $\Xi_{3/2}(1860)$,
    when their spin and parity are assigned to be $J^P=3/2^-$,
    together with other $J^P=3/2^-$ resonances.
    The decay widths of  $\Theta\to KN$,
    $N(1520)\to \pi N$, and $N(1700)\to \pi N$
    are also reproduced simultaneously.
\end{abstract}

\maketitle

%%%%%%%%%%%%%%%%%%%%%%%%%%%%%%%%%%%%%%%%%%%%%%%%%%%%%%%%%%%%%%%%%%%%%%
\section{Introduction}\label{sec:intro}
%%%%%%%%%%%%%%%%%%%%%%%%%%%%%%%%%%%%%%%%%%%%%%%%%%%%%%%%%%%%%%%%%%%%%%

In recent years, there has been a remarkable development
in hadron spectroscopy.
One of the most interesting observations is
the evidence of the exotic baryon $\Theta^+$, reported first
by the LEPS collaboration~\cite{Nakano:2003qx}.
Subsequently, a signal of another exotic state $\Xi^{--}$
was observed~\cite{Alt:2003vb}.
The spin and parity  of $\Theta^+$ and $\Xi^{--}$
are not yet determined experimentally.
Since these states can be constructed minimally with five valence quarks,
they are called pentaquarks.
Evidences of the exotic pentaquarks have
been stimulating many theoretical studies~\cite{Roy:2003hk,Zhu:2004xa,
Oka:2004xh,Diakonov:2004ie,Jaffe:2004ph,Sasaki:2004vz,Goeke:2004ht}.

In the study of these exotic particles,
it should be important to identify other members with nonexotic
flavors in the same SU(3) multiplet which the exotic particles
belong to.
This is naturally expected from the successes of $SU(3)$ flavor
symmetry with its breaking in hadron masses and 
interactions~\cite{deSwart:1963gc}.
In other words, the existence of exotic particles
would require the flavor partners,
if the flavor SU(3) symmetry
plays the same role as in the ordinary
three-quark baryons.

An interesting proposal was made by Jaffe and Wilczek~\cite{Jaffe:2003sg},
based on the assumption of the strong diquark correlation in hadrons
and the representation mixing of an octet ($\bm{8}$) with 
an antidecuplet ($\overline{\bm{10}}$).
The attractive diquark correlation in the scalar-isoscalar channel
leads to the spin and parity $J^P=1/2^+$ 
for the $\Theta^+$.
With the ideal mixing of $\bm{8}$ and $\overline{\bm{10}}$,
in which states are classified by the number  of strange and
antistrange quarks, $N(1710)$ and 
$N(1440)$ resonances are well fit as members of the multiplet
together with the $\Theta^+$.
However, it was pointed out that mixing angles close to the ideal one
encountered a problem in the decay pattern
of $N(1710)\to \pi N$ and $N(1440)\to \pi N$.
Rather, their decays implied a small mixing 
angle~\cite{Cohen:2004gu,Pakvasa:2004pg,Praszalowicz:2004xh}.
This is intuitively understood
by observing the broad decay width of $N(1440)\to \pi N$
and the narrow widths of $N(1710)\to \pi N$ and 
$\Theta\to K N$~\cite{Cohen:2004gu}.

Employing the $\bm{8}$-$\overline{\bm{10}}$ mixing scenario,
here we examine the possibilities to assign other quantum
numbers, such as $1/2^-$, $3/2^+$, $3/2^-$, and search 
the nucleon partners among the known resonances.  
For convenience, 
properties of relevant resonances
are summarized in Appendix~\ref{sec:Expinfo}.

The present study is based on the
flavor SU(3) symmetry, experimental
mass spectra and decay widths of the $\Theta^+$, the $\Xi^{--}$
and known baryon resonances.
Hence, our analysis presented here is phenomenological, but
does not rely upon any specific models.
For instance, we do not have to specify the quark contents of the baryons.
Although the exotic states require minimally five quarks, nonexotic 
partners do not have to. 
Instead, we expect that the resulting properties such as masses and 
decay rates reflect information from which we hope to learn internal
structure of the baryons.

%%%%%%%%%%%%%%%%%%%%%%%%%%%%%%%%%%%%%%%%%%%%%%%%%%%%%%%%%%%%%%%%%%%%%%%%
\section{Analysis with pure antidecuplet}\label{sec:bar10}
%%%%%%%%%%%%%%%%%%%%%%%%%%%%%%%%%%%%%%%%%%%%%%%%%%%%%%%%%%%%%%%%%%%%%%%%

First we briefly discuss the case where the
$\Theta^+$ belongs to the pure $\overline{\bm{10}}$
without mixing with other representations.
In this case, the masses of particles belonging to the $\overline{\bm{10}}$ 
can be determined by the Gell-Mann--Okubo (GMO) mass formula with equal  
splitting
\begin{align}
    M(\overline{\bm{10}};Y)
    &\equiv \bra{\overline{\bm{10}};Y}\mathcal{H}
    \ket{\overline{\bm{10}};Y}
    =  M_{\overline{\bm{10}}} - aY  ,
    \label{eq:bar10mass_1}
\end{align}
where $Y$ is the hypercharge of the state,
and $\mathcal{H}$ denotes the mass matrix.
Note that at this point the spin and parity $J^P$ are not yet
specified. This will be assigned as explained below.

In Eq.~\eqref{eq:bar10mass_1},
there are two parameters, $M_{\overline{\bm{10}}}$ and $a$, which are not
determined by the flavor SU(3) symmetry.
However, we can estimate the order of these parameters
by considering their physical meanings.
For instance, in a constituent quark model,
$\overline{\bm{10}}$ can be minimally expressed as four 
quarks and one antiquark. Therefore, $M_{\overline{\bm{10}}}$
should be larger than the masses of three-quark 
baryons, such as the lowest-lying octet baryons.
In this picture, the
mass difference of $\Xi(ssqq\overline{q})$ and $\Theta(qqqq\overline{s})$,
namely $3a$, should
be the constituent mass difference of the $s$ and the $ud$ quarks,
which is about 100-250 MeV~\cite{Hosaka:2004mv}.
On the other hand, in the chiral quark soliton model,
$3a$ is related to the pion nucleon sigma term~\cite{Schweitzer:2003fg}.
In this picture $3a$ can take values in the range of 300-400 MeV,
due to the experimental uncertainty of the pion nucleon sigma 
term $\Sigma_{\pi N}=$64-79 MeV~\cite{Diakonov:2003jj,
Praszalowicz:2004mt,Ellis:2004uz}.
Note that in the chiral quark model, spin and parity
are assigned as $J^P=1/2^+$ for the antidecuplet.

Taking into account the above estimation, we test several
parameter sets fixed by the experimentally known masses
of particles.
The results are summarized 
in Table~\ref{tbl:bar10result}.
First, we determine the parameters by accommodating $\Theta(1540)$
and $\Xi(1860)$ in the multiplet.
In this case we obtain the mass of the $N$ and $\Sigma$ states at
1647 and 1753 MeV, respectively. 
Since these values are close to the masses
of the $1/2^-$ baryons $N(1650)$ and $\Sigma(1750)$,
we expect their spin and parity to be $J^P=1/2^-$. 
For $J^P=1/2^+$, we adopt the $N(1710)$ as the nucleon partner,
and predict the $\Sigma$ and $\Xi$ states.
This assignment corresponds to the original
assignment of the prediction~\cite{Diakonov:1997mm}.
For $J^P=3/2^+$, we pick up $N(1720)$, and for $J^P=3/2^-$, N(1700).
In the three cases of $J^P=1/2^+, 3/2^{\pm}$,
the exotic $\Xi$ resonance is predicted to be higher than 2 GeV,
and the inclusion of $\Xi(1860)$ in the same multiplet seems to be 
difficult.
Furthermore, the $\Sigma$ states around 1.8-1.9 GeV are not well
assigned (either two-star for $J^P=1/2^+$, or not seen for
$J^P=3/2^{\pm}$).
Therefore, fitting the masses in the pure antidecuplet scheme seems to favor
$J^P=1/2^-$.

Next we study the decay width of the $N^*$ resonances
with the above assignments.
For the decay of a resonance $R$,
we define the dimensionless coupling constant $g_R$ by
\begin{equation}
    \Gamma_R\equiv 
    g_R^{2}F_I \frac{p^{2l+1}}{M_R^{2l}} ,
    \label{eq:coupling}
\end{equation}
where $p$ is the relative three momentum of the decaying particles
in the resonance rest frame, $\Gamma_R$ and $M_R$ are
the decay width and the mass of the resonance $R$.
$F_I$ is the isospin factor, which takes the value
2 for $\Theta\to KN$ and 3 for $N^*\to \pi N$.
Assuming flavor $SU(3)$ symmetry,
a relation between the coupling constants of $\Theta \to K N$ and 
$N^*\to \pi N$ is given by:
\begin{equation}
    g_{\Theta KN}=\sqrt{6}g_{N^*\pi N}  .
    \label{eq:relation}
\end{equation}
Here we adopt the definition of the coupling constant
in Ref.~\cite{Lee:2004bs}. 
Note that this definition
is different from 
Refs.~\cite{Cohen:2004gu,Pakvasa:2004pg},
in which  $g\equiv \sqrt{g_R^2 F_I}$ is  used.
With these formulae~\eqref{eq:coupling} and \eqref{eq:relation},
we calculate the decay width of the $\Theta^+$
from those of $N^*\to \pi N$ of the nucleon partner.
Results are also shown in Table~\ref{tbl:bar10result}.
We quote the errors coming from experimental uncertainties
in the total decay widths and branching ratios,
taken from the Particle Data Group~\cite{Eidelman:2004wy}.
It is easily seen 
that as the partial wave of the two-body final states becomes higher,
the decay width of the resonance becomes narrower, due to the
effect of the centrifugal barrier.
Considering the experimental width of the $\Theta^+$,
the results of $J^P=3/2^-$, $3/2^+$, $1/2^+$ are acceptable,
but the result of the $J^P=1/2^-$ case, which is of the order of
hundred MeV, is unrealistic.

In summary, it seems difficult to regard the $\Theta^+$ as a member of
the pure antidecuplet $\overline{\bm{10}}$ together with known
resonances of $J^P=1/2^{\pm},3/2^{\pm}$,
in fitting both their masses and decay widths.

\begin{table}[tbp]
    \centering
    \caption{Summary of section~\ref{sec:bar10}. 
    Mass spectra and $\Theta^+$ decay width are shown
    for several assignments of quantum numbers.
    For $1/2^-$
    the masses of $\Theta$
    and $\Xi$ are the input parameters, while for $1/2^+,3/2^{\pm}$,
    the masses of $\Theta$ and $N$ are the input parameters.
    Values in parenthesis are the predictions,
    and we show the candidates to be assigned for the states.
    All values are listed in units of MeV.}
    \begin{ruledtabular}
    \begin{tabular}{cccccl}
	$J^P$ & $M_{\Theta}$ & $M_{N}$ & $M_{\Sigma}$ & $M_{\Xi}$
	& $\Gamma_{\Theta}$ \\
	\hline
	& 1540 & [1647] & [1753] & 1860   &    \\
	$1/2^-$ & & $N(1650)$ & $\Sigma(1750)$ &  & 
	$156.1 \ {}^{+90.8}_{-73.3}$  \\
	&  1540  & 1710  & [1880] &[2050] & \\
	$1/2^+$ & &  & $\Sigma(1880)$ & $\Xi(2030)$ &
	  $\phantom{00}7.2 \ {}^{+15.3}_{-4.6}$  \\
	& 1540  & 1720  & [1900] & [2080] &   \\
	$3/2^+$  &  &  & - & - & $\phantom{0}10.6 \ {}^{+7.0}_{-5.0}$  \\
	& 1540 & 1700  & [1860] & [2020] &   \\
	$3/2^-$ & &  & - & $\Xi(2030)$ 
	 & $\phantom{00}1.3 \ {}^{+1.2}_{-0.9}$  \\
    \end{tabular}
    \end{ruledtabular}
    \label{tbl:bar10result}
\end{table}

%%%%%%%%%%%%%%%%%%%%%%%%%%%%%%%%%%%%%%%%%%%%%%%%%%%%%%%%%%%%%%%%%%
\section{Analysis with octet-antidecuplet mixing}\label{sec:8bar10}
%%%%%%%%%%%%%%%%%%%%%%%%%%%%%%%%%%%%%%%%%%%%%%%%%%%%%%%%%%%%%%%%%%%

In this section we consider the representation mixing
between $\overline{\bm{10}}$ and $\bm{8}$. 
In principle, it is possible to take into account the mixing with
multiplets of higher dimension, such as $\bm{27}$ and $\bm{35}$.
However, particles in 
such higher representations will have heavier masses. 
Furthermore, the higher representations bring more states
with exotic quantum numbers, which are not controlled by the known
experimental information.
Here we work under the  assumption of minimal
$\bm{8}$-$\overline{\bm{10}}$ mixing.
Also we do not consider the possible mixing with other octets,
such as ground states~\cite{Guzey:2005mc}.

The nucleon and $\Sigma$ states in the $\bm{8}$
will mix with the states
in the $\overline{\bm{10}}$ of the same quantum numbers.
Denoting the mixing angles of the $N$ and the $\Sigma$ as
$\theta_N$ and $\theta_{\Sigma}$,
the physical states are represented as
\begin{equation}
    \begin{split}
	 \ket{N_1} =& \ket{\bm{8},N} \cos\theta_N
	 - \ket{\overline{\bm{10}},N} \sin\theta_N  , \\
	 \ket{N_2} =& \ket{\overline{\bm{10}},N} \cos\theta_N
	 + \ket{\bm{8},N} \sin\theta_N  ,
    \end{split}
    \label{eq:Nmixing}
\end{equation}
and 
\begin{equation}
    \begin{split}
	 \ket{\Sigma_1} =& \ket{\bm{8},\Sigma} \cos\theta_\Sigma
	 - \ket{\overline{\bm{10}},\Sigma} \sin\theta_\Sigma  , \\
	 \ket{\Sigma_2} =& \ket{\overline{\bm{10}},\Sigma} \cos\theta_\Sigma
	 + \ket{\bm{8},\Sigma} \sin\theta_\Sigma  .
    \end{split}
    \label{eq:Sigmamixing}
\end{equation}
To avoid redundant duplication,
the domain of the mixing angles is restricted in $0\leq \theta < \pi/2$,
and we will find solutions for  $N_1$ and $\Sigma_1$  lighter 
than $N_2$ and $\Sigma_2$, respectively.
The reason for these restrictions is explained in Appendix~\ref{sec:Mixing}.

When we construct $\overline{\bm{10}}$ and $\bm{8}$ from five quarks,
the eigenvalues of the strange quark (antiquark) number operator $n_s$
of nucleon states become fractional.
In the scenario of the ideal mixing of Jaffe and Wilczek,
the physical states are given as
\begin{align}
    \ket{N_1}
    &= \sqrt{\frac{2}{3}}\ket{\bm{8},N}
    -\sqrt{\frac{1}{3}}\ket{\overline{\bm{10}},N}  ,
    \label{eq:idealN1} \\
    \ket{N_2}
    &= \sqrt{\frac{2}{3}}\ket{\overline{\bm{10}},N}
    +\sqrt{\frac{1}{3}}\ket{\bm{8},N} ,
    \label{eq:idealN2}
\end{align}
such that $\bra{N_1}n_s\ket{N_1}=0$ and $\bra{N_2}n_s\ket{N_2}=2$.
In this case, the mixing angle is
\begin{equation}
    \theta_N\sim 35.2^{\circ}  .
    \label{eq:Nideal}
\end{equation}
This value will be compared with the angle obtained from the mass
spectrum of known resonances.
In the Jaffe-Wilczek model~\cite{Jaffe:2003sg},
N(1440) and N(1710) are assigned
to $N_1$ and $N_2$, respectively.
Notice that the separation of the $s\bar{s}$ component in the ideal
mixing is only meaningful for mixing between five-quark states,
while the number of quarks in the baryons is arbitrary in the
present general framework.

It is worth mentioning that the mixing angle $\theta_N$ 
for $1/2^+$ case is calculated
through the dynamical study of constituent quark 
model~\cite{Stancu:2004du}. The resulting value is 
$\theta_N\sim 35.34^{\circ}$, which is very close to the ideal mixing
angle~\eqref{eq:Nideal}.

\subsection{Mass spectrum}\label{subsec:mass}

Let us start with the 
GMO mass formulae for $\overline{\bm{10}}$ and $\bm{8}$ :
\begin{align}
    M(\overline{\bm{10}};Y)
    &\equiv \bra{\overline{\bm{10}};Y} \mathcal{H} 
    \ket{\overline{\bm{10}};Y}
    =  M_{\overline{\bm{10}}} - aY  ,
    \label{eq:bar10mass} \\
    M(\bm{8};I,Y)
    &\equiv \bra{\bm{8};I,Y} \mathcal{H} \ket{\bm{8};I,Y}
    \nonumber \\
    &= M_{\bm{8}} - bY + c
    \left[ I(I+1) -\frac{1}{4}Y^2\right]  ,
    \label{eq:8mass}
\end{align}
where $Y$ and $I$ are the hypercharge and the isospin of the state.
Under representation mixing as in Eqs.~\eqref{eq:Nmixing} and
\eqref{eq:Sigmamixing},
the two nucleons $(N_{\bm{8}},N_{\overline{\bm{10}}})$
and the two sigma states
$(\Sigma_{\bm{8}},\Sigma_{\overline{\bm{10}}})$
mix, and their mass matrices are given by $2\times 2$ matrices.
The diagonal components are given by Eqs.~\eqref{eq:bar10mass}
and \eqref{eq:8mass}, while the off-diagonal elements are given as
\begin{equation}
    \bra{\bm{8},N}\mathcal{H}\ket{\overline{\bm{10}},N}
    =\bra{\bm{8},\Sigma}\mathcal{H}\ket{\overline{\bm{10}},\Sigma}
   \equiv \delta  .
    \label{eq:delta}
\end{equation}
The equivalence of the two off-diagonal elements can be verified when
the symmetry breaking term is given by $\lambda_8$
due to the large strange quark mass~\cite{Diakonov:2003jj}.

The physical states $\ket{N_i}$ and $\ket{\Sigma_i}$
diagonalize $\mathcal{H}$.
Therefore,
we have the relations
\begin{equation}
    \tan 2\theta_N
    = \frac{2\delta}{M_{\overline{\bm{10}}}
    -M_{\bm{8}}-a+b-\frac{1}{2}c}, \label{eq:Nmix}
\end{equation}
and 
\begin{equation}
    \tan 2\theta_{\Sigma}
    = \frac{2\delta}{M_{\overline{\bm{10}}}
    -M_{\bm{8}}-2c}.  \label{eq:Sigmamix}
\end{equation}
Now we have the mass formulae for the states
\begin{align}
    M_{\Theta}
    =& M_{\overline{\bm{10}}}-2a  ,
    \label{eq:Mtheta} \\
    M_{N_1}
    =& \left(M_{\bm{8}}-b+\frac{1}{2}c\right)\cos^2\theta_N
    +\left(M_{\overline{\bm{10}}}-a\right)\sin^2\theta_N 
    \nonumber \\
    &-\delta \sin 2\theta_N ,
    \label{eq:MN1} \\
    M_{N_2}
    =& \left(M_{\bm{8}}-b+\frac{1}{2}c\right)\sin^2\theta_N
    +\left(M_{\overline{\bm{10}}}-a\right)\cos^2\theta_N
    \nonumber \\
    &+\delta \sin 2\theta_N   ,  
    \label{eq:MN2} \\
    M_{\Sigma_1}
    =& \left(M_{\bm{8}}+2c\right)\cos^2\theta_{\Sigma}
    +M_{\overline{\bm{10}}}\sin^2\theta_{\Sigma}
    -\delta \sin 2\theta_{\Sigma} ,
    \label{eq:MSigma1} \\
    M_{\Sigma_2}
    =& \left(M_{\bm{8}}+2c\right)\sin^2\theta_{\Sigma}
    +M_{\overline{\bm{10}}}\cos^2\theta_{\Sigma}
    +\delta \sin 2\theta_{\Sigma} ,
    \label{eq:MSigma2} \\
    M_{\Lambda}
    =&M_{\bm{8}} ,
    \label{eq:MLambda} \\
    M_{\Xi_8}
    =&M_{\bm{8}}+b+\frac{1}{2}c ,
    \label{eq:MXi8} \\
    M_{\Xi_{\overline{\bm{10}}}}
    =&M_{\overline{\bm{10}}}+a .
    \label{eq:MXibar10}
\end{align}
We have altogether six parameters $M_{\bm{8}}$,
$M_{\overline{\bm{10}}}$, $a$, $b$, $c$ and $\delta$.

\begin{figure*}[tbp]
    \centering
    \includegraphics[width=16cm,clip]{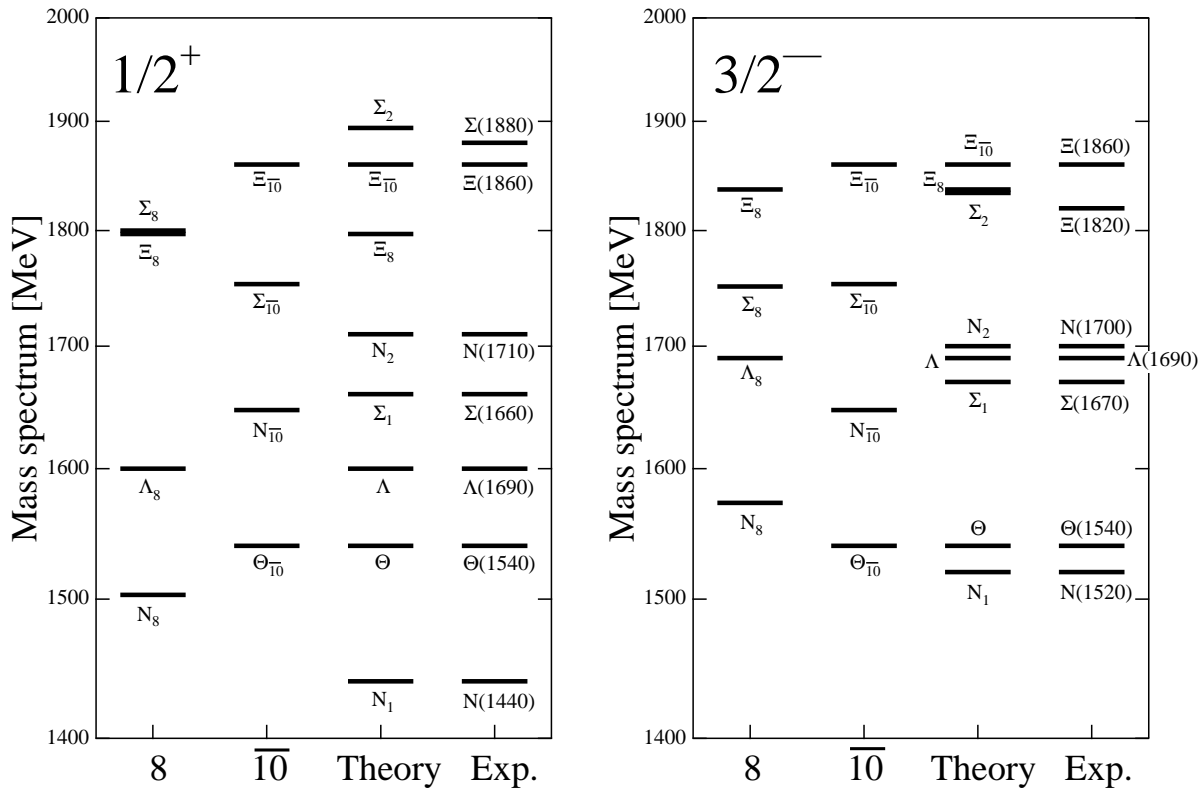}
    \caption{Results of mass spectra with representation mixing.
    Theoretical masses of the octet, antidecuplet, and the one with mixing
    are compared with the experimental masses.
    In the left panel, we show the results with $J^P=1/2^+$,
    while the results with $J^P=3/2^-$ 
    (set 1) are presented in the right panel.}
    \label{fig:spectrum}
\end{figure*}

Let us first examine the case of $J^P=1/2^+$~\cite{Pakvasa:2004pg}.
Possible candidates for the partners of the exotic states
$\Theta(1540)$ and $\Xi_{\overline{10}}(1860)$ are the following:
\begin{align}
    &N(1440), \ N(1710),
    \nonumber \\
    &\Lambda(1600),
    \nonumber \\
    &\Sigma(1660),\ \Sigma(1880).
    \nonumber 
\end{align}
In order to fix the six parameters, we need 
to assign six particles as input.
Using 
$\Theta(1540)$, $N_1(1440)$, $N_2(1710)$,
$\Lambda(1600)$, $\Sigma_1(1660)$, $\Xi_{\overline{10}}(1860)$,
we obtain the parameters as given in Table~\ref{tbl:param1}.
The resulting mass spectrum together with the two 
predicted masses,
$\Sigma_1=1894$ MeV and $\Xi_8=1797$ MeV, are given in
Table~\ref{tbl:result1}
and also shown in the left panel of Fig.~\ref{fig:spectrum}.
For reference, in Table~\ref{tbl:param1} and 
\ref{tbl:result1} we show the parameters and masses of
Ref.~\cite{Pakvasa:2004pg}, in which 
all known resonances including $\Sigma(1660)$ and $\Sigma(1880)$
are used to perform the $\chi^2$ fitting.
In Fig.~\ref{fig:spectrum}, the spectra from experiment
and those before the representation
mixing are also plotted.

As we see in Table~\ref{tbl:result1} and Fig.~\ref{fig:spectrum},
without using the $\Sigma_2$ for the fitting,
this state appears in the proper position to be assigned as
$\Sigma(1880)$.
Considering the experimental uncertainty in the masses, 
these two parameter sets (the one
determined in this work and the one
in Ref.~\cite{Pakvasa:2004pg})
can be regarded as the same one.
In both cases, we need a new $\Xi$ state
around 1790-1800 MeV,
but the overall description of the mass spectrum
is acceptable.
Note that the mixing angle $\theta_N\sim 30^{\circ}$
is compatible with the one of the ideal mixing~\eqref{eq:Nideal},
if we consider the experimental uncertainty
of masses~\cite{Pakvasa:2004pg}.

It is interesting to observe that in the spectrum of the octet,
as shown in Fig.~\ref{fig:spectrum},
the $\Xi_8$ and the $\Sigma_8$ are almost degenerate,
reflecting the large value for the parameter $c\sim$ 100 MeV,
which is responsible for the splitting of $\Lambda$ and $\Sigma$.
For the ground state octet, Eq.~\eqref{eq:8mass}
is well satisfied with $b=139.3$ MeV and $c=40.2$
MeV~\cite{Diakonov:2003jj}. 
This point will be discussed later.

\begin{table}[tbp]
    \centering
    \caption{Parameters for $1/2^+$ case.
    All values are listed in MeV except for the mixing angles.}
    \begin{ruledtabular}
    \begin{tabular}{lllllllll}
	 & $M_{\bm{8}}$ & $M_{\overline{\bm{10}}}$ & $a$
	 & $b$ & $c$ & $\delta$
	 & $\theta_N$ & $\theta_{\Sigma}$  \\
	\hline
	This work & 1600 & 1753.3 & 106.7 & 146.7 & 100.1 & 114.4
	& $29.0^{\circ}$ & $50.8^{\circ}$ \\
	Ref.~\cite{Pakvasa:2004pg}
	& 1600 & 1755 & 107 & 144 & 93 & 123
	& $29.7^{\circ}$ & $41.4^{\circ}$ \\
    \end{tabular}
    \end{ruledtabular}
    \label{tbl:param1}
\end{table}

\begin{table}[tbp]
    \centering
    \caption{Mass spectra for $1/2^+$ case. All values are listed in
    MeV.
    Values in parenthesis ($\Sigma_2$ and $\Xi_{\bm{8}}$
    of Set 1, $\Xi_8$ of Ref.~\cite{Pakvasa:2004pg})
    are predictions (those which are not used in the fitting).}
    \begin{ruledtabular}
    \begin{tabular}{ccccccccc}
	& $\Theta$ & $N_1$ & $N_2$ & $\Sigma_1$ & $\Sigma_2$
	& $\Lambda$ & $\Xi_{\bm{8}}$ & $\Xi_{\overline{\bm{10}}}$  \\
	\hline
	This work & 1540 & 1440 & 1710 & 1660
	& [1894]  & 1600 & [1797]  & 1860  \\
	Ref.~\cite{Pakvasa:2004pg}
	& 1541 & 1432 & 1718 & 1650
	&  1891 & 1600 & [1791]   & 1862 
    \end{tabular}
\end{ruledtabular}
    \label{tbl:result1}
\end{table}

%--figure---------------------------------
%--figure---------------------------------

Now we examine the other cases of $J^P$.
For $J^P={1/2^-}$, as we observed in the previous 
section, the pure $\overline{\bm{10}}$ assignment works well
for the mass spectrum, which implies that the mixing
with $\bm{8}$  is small,
as long as we adopt $N(1650)$ and $\Sigma(1750)$ in the multiplet.
Then the results of $1/2^-$ with the mixing 
do not change from the previous results of the pure 
$\overline{\bm{10}}$  assignment, 
which eventually lead to
a broad width of $\Theta^+\to KN$ of order 100 MeV.
Hence, it is not realistic to assign $1/2^-$,
even if we consider the representation mixing.

Next we consider the $3/2^+$ case.
In this case candidate states
are not well established.
As we see in Appendix~\ref{sec:Expinfo},
no state exists for $\Sigma$ and $\Xi$, except for
two- or one-star resonances.
Furthermore, the states are distributed
in a wide energy range,
and sometimes
it is not possible to assign these
particles in the $\bm{8}$-$\overline{\bm{10}}$
representation scheme.
For instance, if we choose
N(1720), N(1900), $\Lambda(1890)$, $\Sigma(1840)$
and exotic states, no solution is found for the mixing angle.
Therefore, at this moment, it is not meaningful to study
the $3/2^+$ case unless more states 
with $3/2^+$ will be observed.

Now we look at the $3/2^-$ case.
In contrast to the $3/2^+$ case, 
there are several well-established resonances.
Possible candidates are
\begin{align*}
    &N(1520),\  N(1700), \\
    &\Lambda(1520),\  \Lambda(1690), \\
    &\Sigma(1670),\  \Sigma(1940), \\
    &\Xi(1820).
\end{align*}
Following the same procedure as before,
we first 
choose the following four resonances as inputs: 
$\Theta(1540)$, $N_1(1700)$, $N_2(1520)$,
and $\Xi_{3/2}(1860)$.  
For the remaining two to determine the six parameters, 
we examine four different choices of $\Sigma$ and $\Lambda$ states;
\begin{eqnarray}
\Sigma(1670) &{\rm and}& \Lambda(1690)\; \; \;  {\rm (set 1)}, 
\nonumber \\
\Sigma(1670) &{\rm and}& \Lambda(1520)\; \; \;  {\rm (set 2)},
\nonumber \\
\Sigma(1940) &{\rm and}& \Lambda(1690)\; \; \;  {\rm (set 3)}, 
\nonumber \\
\Sigma(1940) &{\rm and}& \Lambda(1520)\; \; \;  {\rm (set 4)}.
\end{eqnarray}
We have obtained the parameters as given in Table~\ref{tbl:param2},
and predicted masses of other members are shown in
Table~\ref{tbl:result2}.
The masses of $N(1520)$ and $N(1700)$ determine the 
mixing angle of nucleons $\theta_N\sim 33^{\circ}$,
which is close to the ideal one.
In the parameter sets 1 and 2 (sets 3 and 4), 
the $\Sigma(1670)$ state of a lower mass 
(the $\Sigma(1940)$ state of a higher mass) 
is chosen but with different $\Lambda$'s, 
$\Lambda(1690)$ and $\Lambda(1520)$.
Accordingly, they predict the higher 
$\Sigma(1834)$ state (the lower $\Sigma(1717)$ state) 
with the mixing angle 
$\theta_\Sigma = 44.6^{\circ} (= 66.2^{\circ})$.  
Interestingly,
parameters of set 1 provide $M_{\Xi_8}\sim 1837$ MeV,
which is close to the known three-star resonance $\Xi(1820)$
of $J^P=3/2^-$.
Parameters of set 4 predict  $M_{\Xi_8}\sim 1659$ MeV,
which is close to another known resonance $\Xi(1690)$.
Since the $J^P$ of this state is not known, this fitting scheme
predicts $J^P$ of $\Xi(1690)$ to be $3/2^-$.
In these two cases, 
we have obtained acceptable assignments, especially for
set 1, although a
new $\Sigma$ state is necessary to complete
the multiplet in both cases.
The spectrum of set 1 is also shown in Fig.~\ref{fig:spectrum}.

Let us briefly look at the octet and antidecuplet 
spectra of $1/2^+$ and $3/2^-$ resonances as shown in 
Fig.~\ref{fig:spectrum}.  
The antidecuplet spectrum is simple, since the GMO
mass formula contains only one parameter which describes the 
size of the splitting.  
Contrarily, the octet spectrum contains two parameters which 
could reflect more information on different internal structure.  
As mentioned before, in the octet spectrum of $1/2^+$, 
the mass of $\Sigma_8$ is pushed up slightly above 
$\Xi_8$, significantly higher than $\Lambda_8$.  
This pattern resembles the octet spectrum which is obtained 
in the Jaffe-Wilczek model, where the baryons are made with 
two flavor $\bar{\bm{3}}$ diquarks and one antiquark.  
In contrast, the spectrum of the octet of $3/2^-$ 
resembles the one of the ground state octet, 
what is reflected in the parameters $(b,c)=(131.9,30.5)$ MeV, 
close to $(b,c)=(139.3,40.2)$ MeV for
the ground states.
This is not far from the prediction of an 
additive quark model of three valence quarks.  
It would be interesting to investigate further the quark 
contents from such a different pattern of the mass spectrum.

\begin{table}[tbp]
    \centering
    \caption{Parameters for $3/2^-$ case.
    All values are listed in MeV except for the mixing angles.}
    \begin{ruledtabular}
    \begin{tabular}{lllllllll}
	 & $M_{\bm{8}}$ & $M_{\overline{\bm{10}}}$ & $a$
	 & $b$ & $c$ & $\delta$
	 & $\theta_N$ & $\theta_{\Sigma}$ \\
	\hline
	set 1 & 1690 & 1753.3 & 106.7 & 131.9 & \phantom{0}30.5 & 82.2
	& $33.0^{\circ}$ & $44.6^{\circ}$ \\
	set 2 & 1520 & 1753.3 & 106.7 & \phantom{00}4.4 & 115.5 & 82.2
	&  $33.0^{\circ}$ & $44.6^{\circ}$ \\
	set 3 & 1690 & 1753.3 & 106.7 & 170.1 & 106.9 & 82.2
	&  $33.0^{\circ}$ & $66.2^{\circ}$ \\
	set 4 & 1520 & 1753.3 & 106.7 & \phantom{0}42.6 & 191.9 & 82.2
	&  $33.0^{\circ}$ & $66.2^{\circ}$ \\
    \end{tabular}
    \end{ruledtabular}
    \label{tbl:param2}
\end{table}

\begin{table}[tbp]
    \centering
    \caption{Mass spectra for $3/2^-$ case. All values are listed in
    MeV.
    Values in parenthesis 
    are predictions (those which are not used in the fitting).}
    \begin{ruledtabular}
    \begin{tabular}{ccccccccc}
	 & $\Theta$ & $N_1$ & $N_2$ & $\Sigma_1$ & $\Sigma_2$
	& $\Lambda$ & $\Xi_{\bm{8}}$ & $\Xi_{\overline{\bm{10}}}$  \\
	\hline
	set 1 & 1540 & 1520 & 1700 & 1670
	& [1834]  & 1690 & [1837]  & 1860  \\
	set 2 & 1540 & 1520 & 1700 & 1670
	& [1834]  & 1520 & [1582]  & 1860  \\
	set 3 & 1540 & 1520 & 1700 
	& [1717] & 1940 & 1690 & [1914]  & 1860  \\
	set 4 & 1540 & 1520 & 1700
	& [1717] & 1940 & 1520 & [1659]  & 1860  \\
    \end{tabular}
    \end{ruledtabular}
    \label{tbl:result2}
\end{table}

\subsection{Decay width}

Here we study the consistency of the mixing angle
obtained from mass spectra and the one obtained from
nucleon decay widths.
Using Eq.~\eqref{eq:relation},
we define a universal coupling constant $g_{\overline{\bm{10}}}$ as
\begin{equation}
    g_{\Theta KN} = \sqrt{6} g_{N_{\overline{\bm{10}}} \pi N}
    \equiv g_{\overline{\bm{10}}}  .
    \label{eq:bar10coupling}
\end{equation}
The coupling constants of the $\pi N$ decay modes
from the $N_{\bm{8}}$, $N_1$,
and $N_2$ are defined as
$g_{N_8}$, $g_{N_1}$, and $g_{N_2}$,
respectively.
The coupling constants of the physical nucleons $N_1$ and $N_2$ are
\begin{align}
    g_{N_1}
    &=g_{N_{\bm{8}}} \cos\theta_N
    -\frac{g_{\overline{\bm{10}}}}{\sqrt{6}}\sin\theta_N
    , \label{eq:N1coupling} \\
    g_{N_2}
    &=\frac{g_{\overline{\bm{10}}}}{\sqrt{6}}
    \cos\theta_N
    +g_{N_{\bm{8}}}\sin\theta_N ,\label{eq:N2coupling}
\end{align}
which are related to the decay widths
through Eq.~\eqref{eq:coupling}.
However, we cannot fix the relative phase
between $g_{N_{\bm{8}}}$ and $g_{\overline{\bm{10}}}$.
Hence, there are two possibilities
of mixing angles both of which
reproduce the same decay widths.
In Refs.~\cite{Pakvasa:2004pg,Praszalowicz:2004xh},
one mixing angle is determined
by neglecting $g_{\overline{\bm{10}}}$ in
Eqs.~\eqref{eq:N1coupling} and \eqref{eq:N2coupling},
which is considered to be small due to the narrow width of $\Theta^+$.
Here we include the effect of $g_{\overline{\bm{10}}}$ explicitly.

Let us examine the two cases, $1/2^+$ and $3/2^-$,
in which we have obtained reasonable mass spectra.
The data for decay widths and branching ratios
to the $\pi N$ channel
of relevant nucleon resonances 
are shown in Table~\ref{tbl:Ndecay}.
Using the mixing angle determined from the mass spectrum and
experimental information of $N^*\to \pi N$ decays,
we obtain the decay width of the $\Theta^+$ 
as shown in Table~\ref{tbl:Thetadecay}.
The widths calculated with the ideal mixing angle are
also presented for reference.
Among the two values,
the former corresponds to the same signs of $g_{N_{\bm{8}}}$
and $g_{\overline{\bm{10}}}$ (phase 1),
while the latter to the opposite signs (phase 2).

\begin{table}[tbp]
    \centering
    \caption{Experimental data for the decay of
    $N^*$ resonances.
    Values in parenthesis are the central values
    quoted in PDG~\cite{Eidelman:2004wy}.}
    \begin{ruledtabular}
    \begin{tabular}{ccrr}
	$J^P$ & Resonance & $\Gamma_{tot}$ [MeV]
	& Fraction ($\Gamma_{\pi N}/\Gamma_{tot}$)   \\
	\hline
	$1/2^+$ & N(1440) & 250-450 (350) & 60-70 (65) \% \\
	 & N(1710) & 50-250 (100) & 10-20 (15) \% \\
	$3/2^-$ & N(1520) & 110-135 (120) & 50-60 (55) \%  \\
	 & N(1700) & 50-150 (100) & 5-15 (10) \%  \\
    \end{tabular}
    \end{ruledtabular}
    \label{tbl:Ndecay}
\end{table}

\begin{table}[tbp]
    \centering
    \caption{Decay width of $\Theta^+$ determined from 
    the nucleon decays and the mixing angle obtained from the mass
    spectra.
    Phase 1 corresponds to the same signs of $g_{N_{\bm{8}}}$
    and $g_{\overline{\bm{10}}}$,
    while phase 2 corresponds to the opposite signs.
    All values are listed in MeV.}
    \begin{ruledtabular}
    \begin{tabular}{ccll}
	$J^P$ & $\theta_N$ & Phase 1
	& Phase 2  \\
	\hline
	$1/2^+$ & $29^{\circ}$ (Mass) & 29.1 & 103.3  \\
	& $35.2^{\circ}$ (Ideal) & 49.3 & 131.8  \\
	$3/2^-$ & $33^{\circ}$ (Mass) & \phantom{0}3.1 & \phantom{0}20.0 \\
	 & $35.2^{\circ}$ (Ideal) & \phantom{0}3.9 &
	\phantom{0}21.3 \\
    \end{tabular}
    \end{ruledtabular}
    \label{tbl:Thetadecay}
\end{table}

For the $1/2^+$ case, the width is about 30 MeV when the mixing angle is 
determined by the mass spectrum, while about 50 MeV for the ideal
mixing angle.
Both values exceed the upper bound of the experimentally observed width.
In contrast, the case $3/2^-$ predicts much narrower widths
of the order of a few MeV both for the two mixing angles,
which are compatible with the experimental upper bound
of the $\Theta^+$ width.

Alternatively, we can determine $\theta_N$
using the experimental decay widths of 
$\Theta \to KN$, $N_1\to \pi N$ and $N_2\to \pi N$.
Here we choose the decay width of $\Theta^+$ as 1 MeV.
Using the central values of the decay widths of
N(1440) and N(1710) and the experimental uncertainty,
we obtain the nucleon mixing angle for the $1/2^+$ case
\begin{equation}
    \begin{split}
	\theta_N =& 6^{\circ} \ {}^{+9^{\circ}}_{-4^{\circ}}  , \\
	\theta_N=&14^{\circ}\ {}^{+10^{\circ}}_{-4^{\circ}}   ,
    \end{split}
    \label{eq:P11}
\end{equation}
where the former corresponds to the phase 1 and 
the latter to the phase 2.
On the other hand,
with $N(1520)$ and $N(1700)$, the mixing angle for the $3/2^-$ case is
\begin{equation}
    \begin{split}
	\theta_N =& 9^{\circ} \ {}^{+9^{\circ}}_{-8^{\circ}}  , \\
	\theta_N=&24^{\circ}\ {}^{+9^{\circ}}_{-9^{\circ}}   .
    \end{split}
    \label{eq:D13}
\end{equation}
For the case of $1/2^+$, the mixing angle of Eq.~\eqref{eq:P11}
may be compared with $\theta_N\sim 30^{\circ}$, which is determined 
from the fitting to the masses. If we consider the large uncertainty 
of the $\pi N$ decay width of $N(1440)$, the mixing
angle~\eqref{eq:P11} can be $24^{\circ}$, which is not very far from
the angle determined by the masses $\theta_N\sim 30^{\circ}$.
On the other hand, for the case of $3/2^-$, the mixing angle~\eqref{eq:D13}
agrees well with the angle determined by the masses $\theta_N\sim 33^{\circ}$.
Considering the agreement of mixing angles and
the relatively small uncertainties in the experimental decay widths,
the results with the $3/2^-$ case are favorable in the present fitting
analysis.

%%%%%%%%%%%%%%%%%%%%%%%%%%%%%%%%%%%%%%%%%%%%%%%%%%%%%%%%%%%%%%%%%%%%%%
\section{Summary and discussion}\label{sec:Summary}
%%%%%%%%%%%%%%%%%%%%%%%%%%%%%%%%%%%%%%%%%%%%%%%%%%%%%%%%%%%%%%%%%%%%%%

We have studied the mass spectra and decay widths of the baryons
belonging to the $\bm{8}$ and $\overline{\bm{10}}$ based on the 
flavor SU(3) symmetry.
As pointed out previously, it was confirmed again
the inconsistency between the mass spectrum
and decay widths of flavor partners
in the octet-antidecuplet mixing scenario
with $J^P=1/2^+$.
However, the assignment of $J^P=3/2^-$ particles
in the mixing  scenario well reproduced the mass spectrum
as well as 
the decay widths of $\Theta(1540)$,
$N(1520)$, and $N(1700)$.
Assignment of $3/2^-$ 
predicted a new $\Sigma$ state
at around 1840 MeV, and the nucleon mixing angle
was close to the one of ideal mixing.
The $1/2^-$ assignment was not realistic
since the widths were too large for $\Theta^+$.
In order to investigate the $3/2^+$ case,
better established experimental data
of the resonances were needed.

The assignment of $J^P=3/2^-$ for exotic baryons seems reasonable
also in a quark model especially when narrow width of the $\Theta^+$
is to be
explained~\cite{Hosaka:2004bn}.
The $(0s)^5$ configuration for the $3/2^-$ $\Theta^+$ is dominated by
the $K^*N$ configuration~\cite{Takeuchi:2004fv}, 
which however cannot be the decay channel
due to the  masses of $K^*$ and $N$ higher than the mass of $\Theta^+$.
Hence we expect naturally (in addition to a naive suppression
mechanism due to the $d$-wave $KN$ decay) a strong suppression of the
decay of the $\Theta^+$.
The possibility
of the spin $3/2$ for the $\Theta^+$ or its excited states
has been discussed not only in quark 
models~\cite{Kanada-Enyo:2004bn,Inoue:2004cc,Hosaka:2004bn,
Takeuchi:2004fv}, but also 
in the $KN$ potential model~\cite{Kahana:2003rv}, 
the $K\Delta$ resonance model~\cite{Sarkar:2004sc}
and QCD sum rule calculations~\cite{Nishikawa:2004tk}.

The $3/2^-$ resonances of nonexotic quantum numbers have been also
studied in various models of hadrons. A conventional quark model
description with a $1p$ excitation of a single quark orbit has been 
successful qualitatively~\cite{Isgur:1978xj}.
Such three-quark states can couple with meson-baryon states which
could be a source for the five- (or more-) quark content of the
resonance.
In the chiral unitary approach, $3/2^-$ states are generated by
$s$-wave scattering states of an octet meson and a decuplet
baryon~\cite{Kolomeitsev:2003kt,Sarkar:2004sc,Sarkar:2004jh}.
By construction, the resulting resonances are largely dominated by
five-quark content. 
These two approaches generate octet baryons which
will eventually mix with the antidecuplet partners to generate the
physical baryons. In other words, careful investigation of the octet 
states before mixing will provide further information.

In the present phenomenological study, we have found that $J^P=3/2^-$
seems to fit observations to date. 
As we have known, other identifications 
have been also discussed in the literature,
for instance, using large $N_c$
expansion~\cite{Jenkins:2004vb,Wessling:2004ag,Pirjol:2004dw}.
It is therefore important to determine the
quantum numbers of $\Theta^+$ in experiments~\cite{Hyodo:2003th,
Nakayama:2003by,Thomas:2003ak,Oh:2003xg,
Hanhart:2003xp,Nam:2004qy,Uzikov:2004bk,Nakayama:2004um,Hanhart:2004re,
Roberts:2004rh,Uzikov:2004er,Oh:2004wp},
not only for the exotic particles
but also for the baryon spectroscopy of nonexotic particles.
Study of high spin states in phenomenological models
and calculations based on QCD are strongly 
encouraged~\cite{Nishikawa:2004tk}.

\begin{acknowledgments}
    We would like to thank Shi-Lin Zhu and Seung-il Nam
    for useful discussions.
    We acknowledge to 
    Manuel J. Vicente Vacas and Daniel Cabrera
    for helpful comments.
    This work supported in part by the Grant for Scientific Research
    ((C) No.16540252) from the Ministry of Education, Culture, 
    Science and Technology, Japan.
\end{acknowledgments}

%%%%%%%%%%%%%%%%%%%%%%%%%%%%%%%%%%%%%%%%%%%%%%%%%%%%%%%%%%%%%%%%%%%%
\appendix
%%%%%%%%%%%%%%%%%%%%%%%%%%%%%%%%%%%%%%%%%%%%%%%%%%%%%%%%%%%%%%%%%%

%%%%%%%%%%%%%%%%%%%%%%%%%%%%%%%%%%%%%%%%%%%%%%%%%%%%%%%%%%%%%%%%%%%%%%%%%
\section{Experimental information}\label{sec:Expinfo}
%%%%%%%%%%%%%%%%%%%%%%%%%%%%%%%%%%%%%%%%%%%%%%%%%%%%%%%%%%%%%%%%%%%%%%%%%

In PDG~\cite{Eidelman:2004wy}, the
masses and widths of $\Theta^+$ and  $\Xi^{--}$ are given as
\begin{align}
    M_{\Theta^+}
    &=1539.2 \pm 1.6 \text{ MeV}  , \quad
    \Gamma_{\Theta^+}
    = 0.9\pm 0.3 \text{ MeV}  .
    \label{eq:Theta} \\
    M_{\Xi^{--}}
    &=1862 \pm 2 \text{ MeV}  , \quad
    \Gamma_{\Xi^{--}}
    < 18 \text{ MeV}  .
    \label{eq:Xi}
\end{align}
In Table~\ref{tbl:resonance} we summarize the resonances with several 
spins and parities.
Note that the $\Sigma(1385)$ and the $\Xi(1530)$ are not listed because 
they are assigned in the decuplet with the $\Delta(1232)$.

%%%%%%%%%%%%%%%%%%%%%%%%%%%%%%%%%%%%%%%%%%%%%%%%%%%%%%%%%%%%%%%%%%%%%%%%%
\section{Mixing angle}\label{sec:Mixing}
%%%%%%%%%%%%%%%%%%%%%%%%%%%%%%%%%%%%%%%%%%%%%%%%%%%%%%%%%%%%%%%%%%%%%%%%%

By looking at the mass formulae given in subsection~\ref{subsec:mass},
the masses of mixed states can be written, in general, by
\begin{align}
    M_1(\theta) &= A \cos^2\theta
    +B \sin^2\theta
    -\frac{(B-A)}{2}\tan 2\theta \sin 2\theta ,
    \label{eq:M1} \\
    M_2(\theta) &= A \sin^2\theta
    +B \cos^2\theta
    +\frac{(B-A)}{2}\tan 2\theta \sin 2\theta .
    \label{eq:M2}
\end{align}
These functions obey the following relations
\begin{align}
    M_i(\theta) =& M_i(\theta + \pi) ,\quad \text{ for } i = 1,2 
    \label{eq:relation1}\\
    M_i(\theta) =& M_i(\pi-\theta) ,\quad \text{ for } i = 1,2 
    \label{eq:relation2}\\
    M_1(\theta) =& M_2(\pi/2-\theta)  , \quad
    M_2(\theta) = M_1(\pi/2-\theta) .
    \label{eq:relation3}	
\end{align}
Equation~\eqref{eq:relation1} shows that 
$M_1(\theta)$ and $M_2(\theta)$ are
periodic functions with period $\pi$,
while
Eq.~\eqref{eq:relation2} shows that
there is a reflection symmetry
of $0\leq\theta\leq\pi/2$
and $\pi/2\leq\theta\leq\pi$.
In order to make a one to one correspondence between
$\theta$ and the masses,
the domain of $\theta$ should be $0\leq\theta<\pi/2$.
In addition, there is a discrete symmetry under the interchange 
$\theta\leftrightarrow\pi/2-\theta$
and $M_1 \leftrightarrow M_2$,
due to Eq.~\eqref{eq:relation3}.
Fixing the assignment of $M_1$ and $M_2$
to the physical states, the mixing angle can be 
determined without duplication.

\begin{table}[htbp]
    \centering
    \caption{Resonances listed in PDG~\cite{Eidelman:2004wy}.
    The $\Sigma(1385)$ and the $\Xi(1530)$ are not listed because 
    they are assigned in the decuplet with the $\Delta(1232)$. 
    We denote stars following the definition in PDG,
    except for three- or four-star resonances
    which are well established.}
    \begin{ruledtabular}
	\begin{tabular}{lll}
	    R & $J^P$ & States  \\
	    \hline
	    $N^*$ & $1/2^-$ & N(1535), N(1650), N(2090)$^*$  \\
	    & $1/2^+$ & N(1440), N(1710), N(2100)$^*$  \\
	    & $3/2^+$ & N(1720), N(1900)$^{**}$  \\
	    & $3/2^-$ & N(1520), N(1700), N(2080)$^{**}$   \\
	    \hline
	    $\Lambda^*$  & $1/2^-$ & $\Lambda(1405)$, $\Lambda(1670)$,
	    $\Lambda(1800)$  \\
	    & $1/2^+$ & $\Lambda(1600)$, $\Lambda(1810)$  \\
	    & $3/2^+$ & $\Lambda(1890)$  \\
	    & $3/2^-$ & $\Lambda(1520)$, $\Lambda(1690)$,
	    $\Lambda(2325)^*$    \\
	     & unknown & $\Lambda(2000)^*$, 
	    $\Lambda(2585)^{**}$    \\
	    \hline
	    $\Sigma^*$  & $1/2^-$ & $\Sigma(1620)^{**}$, $\Sigma(1750)$,
	    $\Sigma(2090)^*$  \\
	    & $1/2^+$ & $\Sigma(1660)$, $\Sigma(1770)^*$,
	    $\Sigma(1880)^{**}$  \\
	    & $3/2^+$ & $\Sigma(1840)^{*}$,
	    $\Sigma(2080)^{**}$  \\
	    & $3/2^-$ & $\Sigma(1580)^{**}$,
	    $\Sigma(1670)$, $\Sigma(1940)$   \\
	    & unknown & $\Sigma(1480)^{*}$, 
	    $\Sigma(1560)^{**}$, $\Sigma(1690)^{**}$, \\
	    &&$\Sigma(2250)$,
	    $\Sigma(2455)^{**}$, $\Sigma(2620)^{**}$\\
	    \hline
	    $\Xi^*$ & $1/2^-$ &   \\
	     & $1/2^+$ &   \\
	     & $3/2^+$ &  \\
	     & $3/2^-$ & $\Xi(1820)$  \\
	    & unknown  & $\Xi(1620)^{*}$, 
	    $\Xi(1690)$, $\Xi(1950)$, $\Xi(2030)$, \\
	    & &
	    $\Xi(2120)^*$, $\Xi(2250)^{**}$, 
	    $\Xi(2370)^{**}$, $\Xi(2500)^*$\\
	\end{tabular}
    \end{ruledtabular}
    \label{tbl:resonance}
\end{table}

% \bibliographystyle{h-physrev3}
% % 
% \bibliography{refs05,refs00,refs90,refs80,refs70,refs60,myrefs,bookrefs}

\end{document}